\documentclass[prd,superscriptaddress,showpacs,nofootinbib,amsmath,amssymb,aps,10pt]{revtex4}

\usepackage{bm}
\usepackage{amsfonts}
\usepackage{latexsym}
\usepackage[latin1]{inputenc}
\usepackage{graphicx}
\usepackage{amsmath}
\usepackage{palatino}
\usepackage{mathpazo}
\usepackage{booktabs}
\usepackage{dcolumn}

\usepackage{wrapfig}
\usepackage{wasysym}
\usepackage[outdir=./]{epstopdf}

\def\jnl@style{\it}
\def\aaref@jnl#1{{\jnl@style#1}}

\def\aaref@jnl#1{{\jnl@style#1}}

\def\aj{\aaref@jnl{AJ}}                   % Astronomical Journal
\def\apj{\aaref@jnl{ApJ}}                 % Astrophysical Journal
\def\apjl{\aaref@jnl{ApJ}}                % Astrophysical Journal, Letters
\def\apjs{\aaref@jnl{ApJS}}               % Astrophysical Journal, Supplement
\def\apss{\aaref@jnl{Ap\&SS}}             % Astrophysics and Space Science
\def\aap{\aaref@jnl{A\&A}}                % Astronomy and Astrophysics
\def\aapr{\aaref@jnl{A\&A~Rev.}}          % Astronomy and Astrophysics Reviews
\def\aaps{\aaref@jnl{A\&AS}}              % Astronomy and Astrophysics, Supplement
\def\mnras{\aaref@jnl{Mon.~Not.~Roy.~Astron.~Soc.}}             % Monthly Notices of the RAS
\def\prd{\aaref@jnl{Phys.~Rev.~D}}        % Physical Review D
\def\prc{\aaref@jnl{Phys.~Rev.~C}}  % Physical Review C
\def\prl{\aaref@jnl{Phys.~Rev.~Lett.}}    % Physical Review Letters
\def\qjras{\aaref@jnl{QJRAS}}             % Quarterly Journal of the RAS
\def\skytel{\aaref@jnl{S\&T}}             % Sky and Telescope
\def\ssr{\aaref@jnl{Space~Sci.~Rev.}}     % Space Science Reviews
\def\zap{\aaref@jnl{ZAp}}                 % Zeitschrift fuer Astrophysik
\def\nat{\aaref@jnl{Nature}}              % Nature
\def\aplett{\aaref@jnl{Astrophys.~Lett.}} % Astrophysics Letters
\def\apspr{\aaref@jnl{Astrophys.~Space~Phys.~Res.}} % Astrophysics Space Physics Research
\def\physrep{\aaref@jnl{Phys.~Rep.}}      % Physics Reports
\def\physscr{\aaref@jnl{Phys.~Scr}}       % Physica Scripta
\def\commat{\aaref@jnl{Comm.~Math.~Phys.}}              % Communications in Mathematical Physics
\def\science{\aaref@jnl{Science}}               % Science
\def\cqg{\aaref@jnl{Classical Quant.~Grav.}}            % Classical and Quantum Gravity
\def\jpcs{\aaref@jnl{JPCS}}                                     % Journal of Physics Conference Series
\def\ijmpd{\aaref@jnl{Int.~J.~Mod.~Phys.~D}}                    % International Journal of Modern Physics D
\def\grg{\aaref@jnl{Gen.~Relat.~Gravit.}}               % General Relativity and Gravitation
\def\rpp{\aaref@jnl{Rep.~Prog.~Phys.}}          % Reports on Progress in Physics
\def\npa{\aaref@jnl{Nucl.~Phys.~A}}        % Nuclear Physics A
\def\lrr{\aaref@jnl{Living Rev.~Rel.}}                   % Living reviews in relativity
\def\jcap{\aaref@jnl{J.~Cosmology Astropart.~Phys.}}    % Journal of cosmology and astroparticle physics
\def\rmp{\aaref@jnl{Rev.~Mod.~Phys.}}   %Reviews of modern physics

\allowdisplaybreaks[1]
\begin{document}

\title{Universal I-Q relations for rapidly  rotating neutron and strange stars in scalar-tensor theories}

\author{Daniela D. Doneva}
\email{daniela.doneva@uni-tuebingen.de}
\affiliation{Theoretical Astrophysics, Eberhard Karls University of T\"ubingen, T\"ubingen 72076, Germany}
\affiliation{INRNE - Bulgarian Academy of Sciences, 1784  Sofia, Bulgaria}

\author{Stoytcho S. Yazadjiev}
\email{yazad@phys.uni-sofia.bg}
\affiliation{Department of Theoretical Physics, Faculty of Physics, Sofia University, Sofia 1164, Bulgaria}
\affiliation{Theoretical Astrophysics, Eberhard Karls University of T\"ubingen, T\"ubingen 72076, Germany}

\author{Kalin V. Staykov}
\email{kalin.v.staikov@gmail.com}
\affiliation{Department of Theoretical Physics, Faculty of Physics, Sofia University, Sofia 1164, Bulgaria}
\affiliation{Theoretical Astrophysics, Eberhard Karls University of T\"ubingen, T\"ubingen 72076, Germany}

\author{Kostas D. Kokkotas}
\email{kostas.kokkotas@uni-tuebingen.de}
\affiliation{Theoretical Astrophysics, Eberhard Karls University of T\"ubingen, T\"ubingen 72076, Germany}
\affiliation{Department of Physics, Aristotle University of Thessaloniki, Thessaloniki 54124, Greece}

%%%%%%%%%%%%%%%%%%%%%%%%%%%%%%%%%%%%  DATE  %%%%%%%%%%%%%%%%%%%%%%%%%%%%%%%%%%%%

\begin{abstract}
We study how rapid rotation influences the relation between the normalized moment of inertia $\bar{I}$ and quadrupole moment $\bar{Q}$ for scalarized neutron stars. The questions one has to answer are whether the EOS universality is preserved in this regime and what are the deviations from general relativity. Our results show that the $\bar{I}-\bar{Q}$ relation is nearly EOS independent for scalarized rapidly rotating stars, but the differences with pure Einstein's theory increase compared to the slowly rotating case. In general, smaller negative values of the scalar field coupling parameters $\beta$ lead to larger deviations, but these deviations are below the expected accuracy of the future astrophysical observations if one considers values of $\beta$  in agreement with the current observational constraint. An important remark is that although the normalized $\bar{I}-\bar{Q}$ relation is quite similar for scalar-tensor theories and general relativity, the unnormalized moment of inertia and quadrupole moment can be very different in the two theories. This demonstrates that although the $\bar{I}-\bar{Q}$ relations are potentially very useful for some purposes, they might not serve us well when trying to distinguish between different theories of gravity.
\end{abstract}

\pacs{}
\maketitle
\date{}
\section{Introduction}
Neutron stars are among the best candidates to test the strong field regime of general relativity (GR) due to the high compactness of the matter in their core and to the variety of observed phenomena. One of the most significant obstacle in this direction is the uncertainty in the nuclear matter equation of state (EOS). In many cases these uncertainties are comparable or even bigger than the effects induced by different modifications of GR. Strong constraints on the EOS can be set \cite{Lattimer12,Steiner2010,Ozel2013,Antoniadis13,Demorest10}, but still the accuracy is below the desired one for testing different generalizations of Einstein's theory. A way to circumvent these uncertainties is to search for equation of state independent characteristics. An important step in this direction was made recently in \cite{Yagi2013,Yagi2013a} (see also \cite{Yagi2014,Yagi2014a,Yagi2014b}), where it was found that relations between the normalized moment of inertia ($I$), quadrupole moment ($Q$) and the tidal Love number ($\lambda$) exist (the so-called $I$-Love-$Q$ relations) which are practically independent of the equation of state. In a subsequent series of papers these results were generalized to large tidal deformations \cite{Maselli2013} and fast rotation \cite{Doneva2014,Pappas2014,Chakrabarti2014}. Magnetized neutron stars on the other hand can break the universality for strong magnetic fields and low rotational rates. In \cite{Yagi2014a} it was shown that similar EOS independent relations can be formulated for the higher multipole moments. Other universal relations were studied in \cite{Lattimer2001,Urbanec2013,Baubock2013,AlGendy2014,Andersson98a,Tsui2005}.

Different astrophysical implications of the $I$-Love-$Q$ relations were proposed. One of the most important is breaking the degeneracy between the spins and the quadrupole moments of neutron star inspirals. A development in this direction is important due to the expected detection of gravitational waves in the very near future. Another possible implication of the $I$-Love-$Q$ relations is to use them as a probe for modified theories of gravity. For example if a specific alternative theory gives a significantly different $I$-Love-$Q$ relations compared to GR, then a measurement of two of these quantities would allow us to determine the possible deviations from general relativity and set constraints on the alternative theory of gravity. This is for example the case with the Dynamical Chern-Simons gravity where the $I$-Love-$Q$ relations are still pretty much EOS independent but they are significantly different from the GR case \cite{Yagi2013,Yagi2013a}. On the other hand the corresponding relations in Eddington-inspired Born-Infeld gravity, scalar-tensor theories of gravity and Einstein-Gauss-Bonnet-dilaton theory \cite{Pani2011b} are almost indistinguishable from Einstein's theory \cite{Sham2014,Pani2014,Kleihaus2014}. Therefore the $I$-Love-$Q$ relations can not serve as a test for these theories. But from another point of view these results show that the relations are universal not only with respect to the EOS but also to certain beyond-GR corrections. Thus if one of the quantities in the $I$-Love-$Q$ trio is measured, one can infer the other two using formulae which are universal for a whole plethora of gravitational theories.

Except the case of Einstein-Gauss-Bonnet-dilaton  gravity \cite{Kleihaus2014}, all of the studies of $I$-Love-$Q$ relations in alternative theories of gravity up to now are limited to the slow rotation regime. One of the reasons is that constructing rapidly rotating neutron star solutions in alternative theories of gravity is very involved. Also the slow rotation approximation is sufficient in practice for many cases, for example the inspiraling neutron stars are supposed to be rotating with relatively low rotational frequencies. But the rotational rates of neutron stars in some cases, such as the millisecond pulsars and newborn neutron stars, can reach very high values and the extension to rapid rotation is important. Moreover the studies of neutron stars in alternative theories of gravity show that rotation can magnify the deviation from GR significantly and potentially lead to observational consequences \cite{Doneva2013}.

In the present paper we concentrate on the $I$-$Q$ relation for rapidly rotating neutron stars in scalar-tensor theory of gravity. This is one of the most natural and widely explored extension of general relativity where in addition to the spacetime metric, a scalar field appears which is also a mediator of the gravitational interaction. Different classes of scalar-tensor theories were considered in the literature with probably the most famous one being the Brans-Dicke theory. But in the past two decades special attention was given to a specific class of scalar-tensor theories which is indistinguishable from GR in the weak field regime but interesting nonlinear effects can develop for strong fields. Such a nonlinear phenomenon is the scalarization of neutron star discovered in \cite{Damour1993}. These results were generalized later to slow \cite{Damour1996,Sotani2012,Pani2014} and rapid rotation \cite{Doneva2013}. The general idea behind the scalarization is that for certain ranges of the parameters new solutions with nontrivial scalar field can exist in addition to the pure GR solutions, and the scalarized neutron stars are energetically more favorable compared to their GR counterpart. The current observations of binary pulsars set tight constraints on the coupling parameter in this scalar-tensor theory and for the nonrotating case the scalarized solutions differ only slightly from the general relativistic ones \cite{Will2006,Freire2012,Antoniadis13}. But as the results in \cite{Doneva2013} show, the rotation can enhance the effect of the scalar field considerably and leads to much more significant deviations from GR.

A natural question that arises is whether the $I$-Love-$Q$ relations are different for scalarized neutron stars. As we have already mentioned, the studies in slow rotation approximation \cite{Pani2014} show that these relations are practically the same as in GR if one considers values of the parameters in agreement with the current observational constrains. Our goal in the current paper is to extend these results to the case of rapid rotation and to check if the universality of the relations is preserved. For this purpose we have to make also a proper derivation of the quadrupole moment of rotating scalarized neutron stars in quasi-isotropic coordinates.

\section{Basic equations}
In this section we very briefly present the necessary background for the scalar-tensor theories and the rapidly rotating stars in their framework without going into too much detail. For  a more detailed discussion we refer the reader to \cite{Doneva2013}.

The scalar-tensor field equations in the Einstein frame are

\begin{eqnarray} \label{EFFE}
R_{\mu\nu} - {1\over 2}g_{\mu\nu}R = 8\pi G_{*} T_{\mu\nu}
 + 2\partial_{\mu}\varphi \partial_{\nu}\varphi   -
g_{\mu\nu}g^{\alpha\beta}\partial_{\alpha}\varphi
\partial_{\beta}\varphi -2V(\varphi)g_{\mu\nu}  \,\,\, ,\nonumber
\end{eqnarray}

\begin{eqnarray}
 \nabla^{\mu}\nabla_{\mu}\varphi = - 4\pi G_{*} k(\varphi)T
+ {dV(\varphi)\over d\varphi} ,
\end{eqnarray}

where $k(\varphi)= {d\ln({\cal  A}(\varphi))/ d\varphi}$  with ${\cal A}(\varphi)$ being the conformal function relating the Jordan and Einstein frame metrics.
In what follows we consider only scalar-tensor theories with $V(\varphi)=0$.  $T_{\mu\nu}$ is the
Einstein frame energy-momentum tensor $T_{\mu\nu}$. In modeling the rotating stars we use the energy-momentum tensor
of a perfect fluid $T_{\mu\nu}=(\epsilon + p)u_{\mu}u_{\nu} + pg_{\mu\nu}$ with energy density $\epsilon$, pressure $p$ and
 4-velocity $u^{\mu}$.

The spacetime metric for rotating stars can be presented in the form

\begin{eqnarray}
ds^2 = - e^{2\nu}dt^2 + \rho^2 B^2 e^{-2\nu}(d\phi - \omega dt)^2 + e^{2\zeta - 2\nu}(d\rho^2 + dz^2),
\end{eqnarray}
where all the metric functions depend on the coordinates $\rho$ and $z$. We will consider the natural from a physical point of view case when the
geometry and the scalar field are invariant under the reflection symmetry through the equatorial plane, i.e. when the metric functions and the scalar field are invariant under
the  map $\theta \to \pi - \theta$.

In comparison with \cite{Doneva2013}, in the resent paper we use a slightly different
but equivalent representation of the metric and the dimensionally reduced field equations.

The dimensionally reduced Einstein frame scalar-tensor
field equations describing the structure of the  rotating stars (with constant angular velocity) are given by
\begin{eqnarray}\label{EFRFE1}
&&B^{-1}D_{i}\left(BD^{i}\nu\right)=\frac{1}{2}\rho^2 B^2 e^{-4\nu} D_{i}\omega D^{i}\omega + 4\pi e^{2\zeta - 2\nu}\left[(\epsilon + p) \frac{1+v^2}{1-v^2} + 2p \right],\\
&&\rho^{-1}D_{i}\left(\rho D^{i}B\right)=16\pi B e^{2\zeta - 2\nu}  p, \\
&&D_{i}D^{i}\omega = \left[4D_{i} - 2\rho^{-1}D_{i}\rho - 3B^{-1}D_{i}B \right] D^{i}\omega - 16\pi \rho^{-1}B^{-1}e^{2\zeta - 2\nu} (\epsilon + p) \frac{v}{1-v^2}, \\
&&\rho^{-1}\partial_{z}\zeta + B^{-1}\left(\partial_{\rho}B\partial_{z}\zeta + \partial_{z}B\partial_{\rho}\zeta\right)
- \frac{1}{2}\rho^{-2}B^{-1} \partial_{\rho}\left(\rho^{2}\partial_{z}B \right) - \frac{1}{2} B^{-1} \partial_{\rho}\partial_{z}B= 2 \partial_{\rho}\nu\partial_{z}\nu
\nonumber \\ && - \frac{1}{2}\rho^{2}B^{2}e^{-4\nu} \partial_{\rho}\omega\partial_{z}\omega    + 2 \partial_{\rho}\varphi\partial_{z}\varphi , \\
&&\rho^{-1}\partial_{\rho}\zeta + B^{-1}\left(\partial_{\rho}B\partial_{\rho}\zeta - \partial_{z}B\partial_{z}\zeta\right)
- \frac{1}{2}\rho^{-2}B^{-1} \partial_{\rho}\left(\rho^{2}\partial_{\rho}B \right) + \frac{1}{2} B^{-1} \partial^2_{z}B= (\partial_{\rho}\nu)^2 - (\partial_{z}\nu)^2
\nonumber \\ && - \frac{1}{4}\rho^{2}B^{2}e^{-4\nu}\left[ (\partial_{\rho}\omega)^2 - (\partial_{z}\omega)^2\right]
+  (\partial_{\rho}\varphi)^2 - (\partial_{z}\varphi)^2, \\
&&B^{-1}D_{i}\left(BD^{i}\varphi\right)=4\pi k(\varphi)\left(\epsilon -3p\right)e^{2\zeta - 2\nu}, \\
&&D_{i}p= - (\epsilon + p)\left(D_{i}\nu - \frac{v}{1-v^2}D_{i}v\right) - k(\varphi)(\epsilon - 3p)D_{i}\varphi
\label{EFRFE7}
\end{eqnarray}
where $D_i$ is the covariant derivative with respect to the 3-dimensional flat metric $dl^2=d\rho^2 + \rho^2 d\phi^2 + dz^2$ and
$v=(\Omega -\omega)B\rho e^{-2\nu}$ is the proper velocity of  the stellar fluid with $\Omega$ being the angular velocity of the star.

In order to derive the formula for the quadrupole moment we will need the system of differential equations that describes the exterior of the star, i.e the  stationary and axisymmetric, vacuum scalar-tensor equations. This system is formally obtained from (\ref{EFRFE1})-(\ref{EFRFE7}) by setting $\epsilon=p=0$. Using these stationary and axisymmetric vacuum equations one can find the asymptotic behaviour
of the metric functions and the scalar field. From a physical point of view it is more convenient and clear to present the asymptotic behaviour in the quasi-isotropic coordinates $r$ and $\theta$ defined by

\begin{eqnarray}
\rho=r\sin\theta , \; \; z=r\cos\theta.
\end{eqnarray}

In these coordinates, keeping only terms up to order of $r^{-3}$ we have

\begin{eqnarray}\label{ASMPT1}
&&\nu \approx - \frac{M}{r} + \left[ \frac{b}{3} + \frac{\nu_2}{M^3} P_{2}(\cos\theta)\right] \left( \frac{M}{r}\right)^3 , \\
&&B\approx 1 + b \left( \frac{M}{r}\right)^2, \\
&&\omega \approx \frac{2J}{r^3} , \\
&& \zeta\approx -\left\{\frac{1}{4}(1 + \frac{{\cal D}^2}{M^2})
+ \frac{1}{3}\left[b + \frac{1}{4}(1 + \frac{{\cal D}^2}{M^2})\right]\left[1 - 4P_{2}(\cos\theta)\right] \right\} \left( \frac{M}{r}\right)^2 ,\\
&&\varphi \approx - \frac{{\cal D}}{r} + \left\{\frac{1}{3}\left(\frac{{\cal D}}{M}\right)b + \frac{\varphi_{2}}{M^3} P_{2}(\cos\theta) \right\} \left( \frac{M}{r}\right)^3
\label{ASMPT2}
\end{eqnarray}
where $M$ and $J$ are the mass and the angular momentum, ${\cal D}$ is the scalar charge defined by
\begin{eqnarray}
{\cal D}= \frac{1}{4\pi} \oint_{S^2_{\infty}} D_i\varphi dS^{i},
\end{eqnarray}
$b$, $\nu_2$ and $\varphi_{2}$ are constants and $P_{2}(\cos\theta)$ is the second Legendre polynomial. In the present paper as in \cite{Doneva2013} we consider the case $\lim_{\infty}\varphi=\varphi_{\infty}=0$.   Proceeding further, we can,  just as in general relativity, derive the formula for the quadrupole moment from the asymptotic expansion of the metric functions. After some algebra we find the following formula for the quadrupole moment in our case

\begin{eqnarray}
Q=- \nu_{2} - \frac{4}{3}\left[b + \frac{1}{4}(1 + \frac{{\cal D}^2}{M^2})\right]M^3.
\end{eqnarray}

As can be seen the quadrupole moment formula explicitly involves the scalar charge and reduces to that in general relativity
for ${\cal D}=0$.

The other quantity we will need in the present paper is the moment of inertia $I$ which is defined as usual, namely

\begin{eqnarray}
I=\frac{J}{\Omega}
\end{eqnarray}
with $\Omega$ being the angular velocity of the star.

The quadrupole moment and the moment of inertia have been defined in the Einstein frame. However, we need these quantities  in the physical Jordan frame. For  the
moment of inertia one  can show that it is the same in both frames \cite{Doneva2013}. The relation between the quadrupole moment in the Einstein and the Jordan frame depends in general on the particular scalar-tensor theory.  In the present paper we consider a class of scalar-tensor theories defined by ${\cal A}^2(\varphi)=e^{\beta\varphi^2}$ (and $V(\varphi)=0$) with $\beta$
being a negative parameter $\beta<0$. For this particular class of scalar-tensor  theories, taking into account the asymptotic expansions of the Einstein frame quantities  given in (\ref{ASMPT1})-(\ref{ASMPT2}) and especially that $\varphi_{\infty}=0$, one can easily find the asymptotic expansion of the Jordan frame metric ${\tilde g}_{\mu\nu}={\cal A}^2(\varphi) g_{\mu\nu}$ which shows that that the quandrupole moment in the Jordan frame ${\tilde Q}$ is the same with that in the Einstein frame, ${\tilde Q}=Q$.

\section{Numerical results}

We calculate the rotating neutron star solutions in scalar-tensor theories of gravity with a modified version of the ${\tt rns}$ code \cite{Stergioulas95} developed in \cite{Doneva2013}. The normalized moment of inertia $\bar{I} = \frac{I}{M^3}$ and the normalized quadrupole moment $\bar{Q} = -\frac{Q}{M^{3}\chi^{2}}$ are examined, where $\chi =\frac{J}{M^2}$. More specifically our goal is to check the universality of the $\bar{I}-\bar{Q}$ relation for rapidly rotating scalarized neutron stars. As we have already commented, the $\bar{I}-\bar{Q}$ relation is practically indistinguishable from GR for slow rotation \cite{Pani2014}, if one considers values of $\beta$ which are in agreement with the current constrains coming from the binary pulsar experiments $\beta>-4.5$ \cite{Will2006,Freire2012,Antoniadis13}.
In our calculations we will use mainly $\beta = -4.5$ which represents the lower limit on $\beta$. In order to demonstrate the sensitivity of our results and conclusions to the value of the coupling parameter $\beta$, we also show some calculations for smaller $\beta$, more precisely $\beta=-5$ and $\beta=-6$
which are already ruled out by the observations.

In order to quantify the deviation from slow rotation we will build sequences of models with constant values of the normalized parameter  $\alpha = fM$, where $f=\Omega/2\pi$ is the rotational frequency of the star, and in our system of units $f$ is given in kHz and $M$ is in solar masses. As it was shown in \cite{Chakrabarti2014,Pappas2014}, this choice of $\alpha$ leads to EOS independent relations in the GR case for each value of $\alpha$. The last bit of information we have to fix is the set of equations of state. We employee six hadronic EOS which span a very wide range of stiffness. These are WFF2 \cite{WFF}, APR \cite{AkmalPR}, GCP \cite{Goriely2010}, HLPS \cite{Hebeler2010}, FPS \cite{Pandharipande81} and the zero temperature limit of Shen EOS \cite{Shen1998,Shen1998a}. For the sake of completeness we consider also two strange star EOS -- the SQSB40 and SQSB60 given in \cite{Gondek-Rosinska2008}.

\begin{figure}[]
\centering
\includegraphics[width=0.48\textwidth]{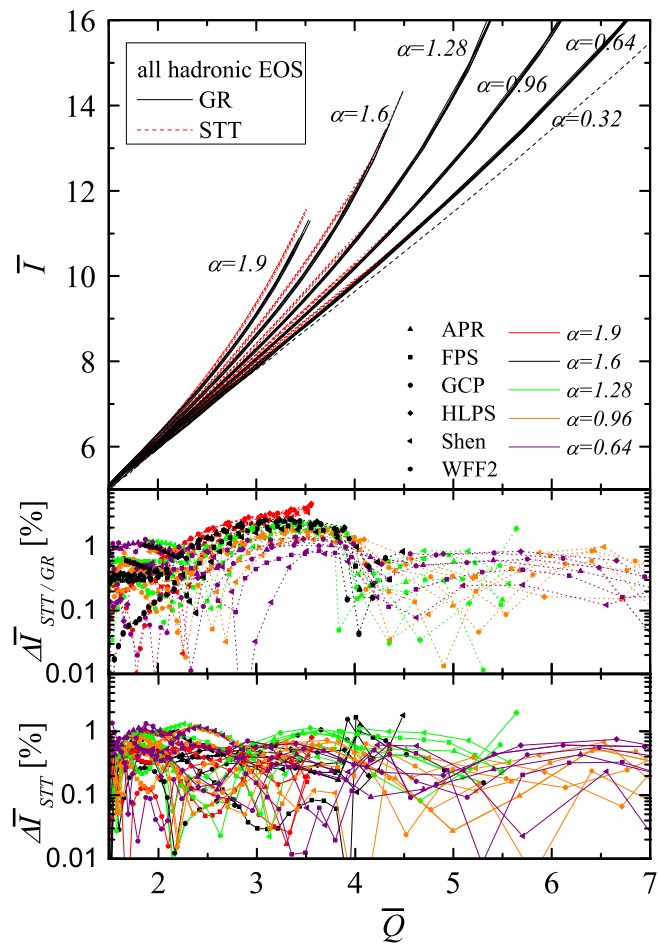}
\caption{$\bar{I}-\bar{Q}$ relations for GR (continuous black lines) and for STT (red lines) for all hadronic EOSs and for several values of the parameter $\alpha$. The case of $\alpha=0.32$ is given as dashed black line as representative for slow rotation (GR and STT solutions in this case are almost indistinguishable). In the middle panel the relative deviation of the STT solutions from the GR polynomial fits is presented. In the bottom panel the relative deviations of STT solutions from the fits to the STT data are plotted. } \label{Fig:I_Q_delta}
\end{figure}

In Fig \ref{Fig:I_Q_delta} we present the $\bar{I} - \bar{Q}$ relation and the associated deviations from universality in the case of scalar-tensor theories with $\beta=-4.5$ and for the pure general relativistic case. The results for all hadronic EOSs are shown. For each value of $\alpha$ we make a fourth order polynomial fit to the input data of the form
\begin{equation}
\ln{\bar{I}}= a_0+a_1\ln{\bar{Q}}+a_2(\ln{\bar{Q}})^2+a_3(\ln{\bar{Q}})^3+a_4(\ln{\bar{Q}})^4.
\end{equation}
The calculated models span the range from slow rotation to the Kepler (mass shedding) limit. We present the results for $\alpha$  up to $1.9$. The slow rotation limit on the other hand, corresponding to $\alpha=0.32$ in the figure, is shown as a dashed line for both GR and STT. The reason is that for slow rotation the two cases are practically indistinguishable when $\beta=-4.5$ and the scalarization can hardly be noticed in the graph.
In Fig. \ref{Fig:I_Q_delta} only the case of hadronic EOS is shown, but our results show that strange stars lead to practically the same $\bar{I} - \bar{Q}$ relation for both the scalarized and non-scalarized cases. The only notable difference is that quark stars can reach higher values of $\bar{Q}$ and $\bar{I}$ for rapid rotation, due to the higher oblateness close to their Kepler limit.

In the two lower panels of Fig. \ref{Fig:I_Q_delta} the relative deviations of the solutions from the polynomial fits are shown, defined as
\begin{equation}
\Delta \bar{I}=\frac{|\bar{I}-\bar{I}_{fit}|}{\bar{I}_{fit}}.
\end{equation}
We plot two deviation -- $\Delta \bar{I}_{STT}$ (bottom panel) which represents the deviation of the data in STT from the calculated STT fits, i.e. the deviation from EOS universality, and $\Delta \bar{I}_{STT/GR}$ (middle panel) which is the deviation of the scalarized neutron stars from the general relativistic fits.

Several conclusions can be made using these results. First it if important to note that for all values of $\alpha$ the deviation from EOS universality for scalarized neutron stars is below approximately $1\%$, similar to the pure GR case. This means that for a fixed value of $\alpha$ the $\bar{I} - \bar{Q}$ relation is indeed nearly EOS independent for scalarized neutron stars. The difference between the $\bar{I} - \bar{Q}$ relations for neutron stars in GR and STT on the other hand increases with rotation, which is an expected results as rapid rotation can significantly magnify the deviations from GR in the neutron star equilibrium properties \cite{Doneva2013}. But still the deviations are below roughly $5\%$ even for the most rapidly rotating models shown on the graph. This difference is above the deviation from EOS universality and it is sufficient to make a clear distinction between the $\bar{I} - \bar{Q}$ relations for scalarized and non-scalarized solutions (at least for rapid rotation). But the deviation is below the expected observational accuracy and therefore it would be difficult to set further constrains on STT using the $\bar{I} - \bar{Q}$ relations.

\begin{figure}[]
\centering
\includegraphics[width=0.6\textwidth]{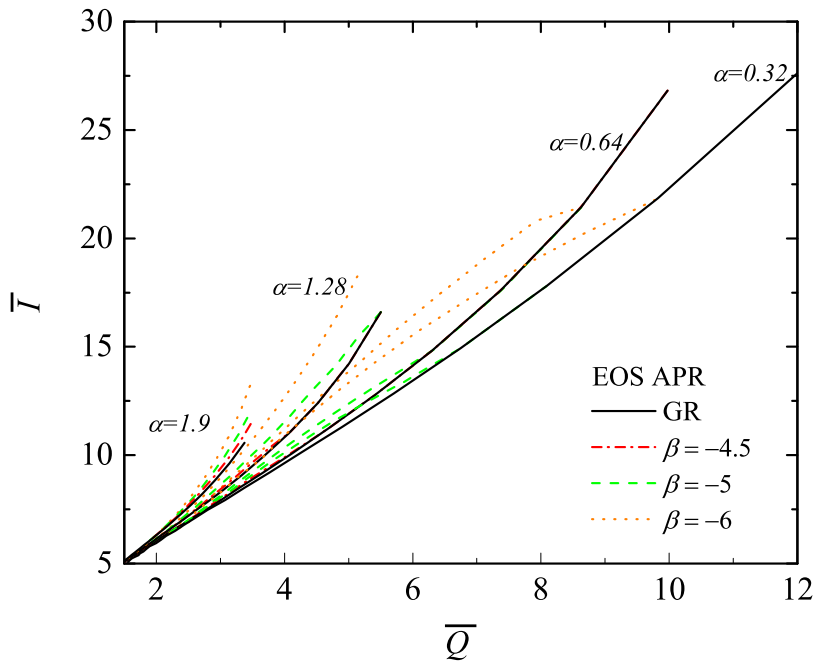}
\caption{Comparison between $\bar{I}-\bar{Q}$ relations for GR and for STT, for several values of $\beta= -4.5, -5, -6$. The presented data are for EOS APR.} \label{Fig:APR_STT_GR}
\end{figure}

Let us now turn to the question of the sensitivity of our results to the value of the scalar field coupling parameter $\beta$. In Fig. \ref{Fig:APR_STT_GR} we examined the behavior of the scalarized solutions for different values of $\beta$ in the case of EOS APR.  We should note that the cases with $\beta=-5$ and $\beta=-6$ are already ruled out by the binary pulsar experiments and they are used just to examine the qualitative behavior. As one can see, when the value of $\beta$ is decreasing, the $\bar{I}-\bar{Q}$ relation can change considerably. This is indeed an expected results because in general smaller negative values of $\beta$ lead to larger deviations from GR, and for example scalarized neutron stars with $\beta=-6$ would already have a very different equilibrium properties even in the nonrotating limit.

\begin{figure}[]
\centering
\includegraphics[width=0.48\textwidth]{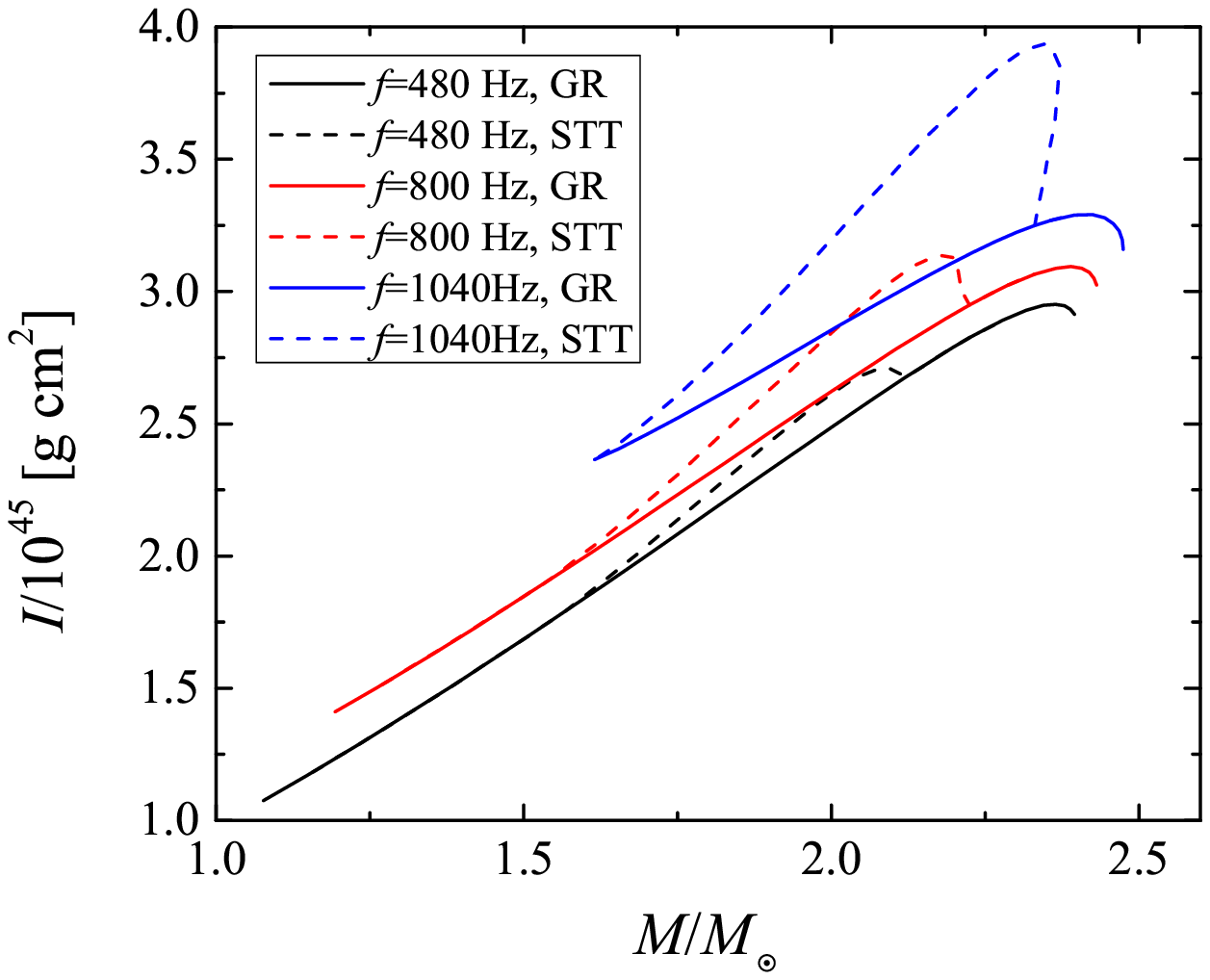}
\includegraphics[width=0.46\textwidth]{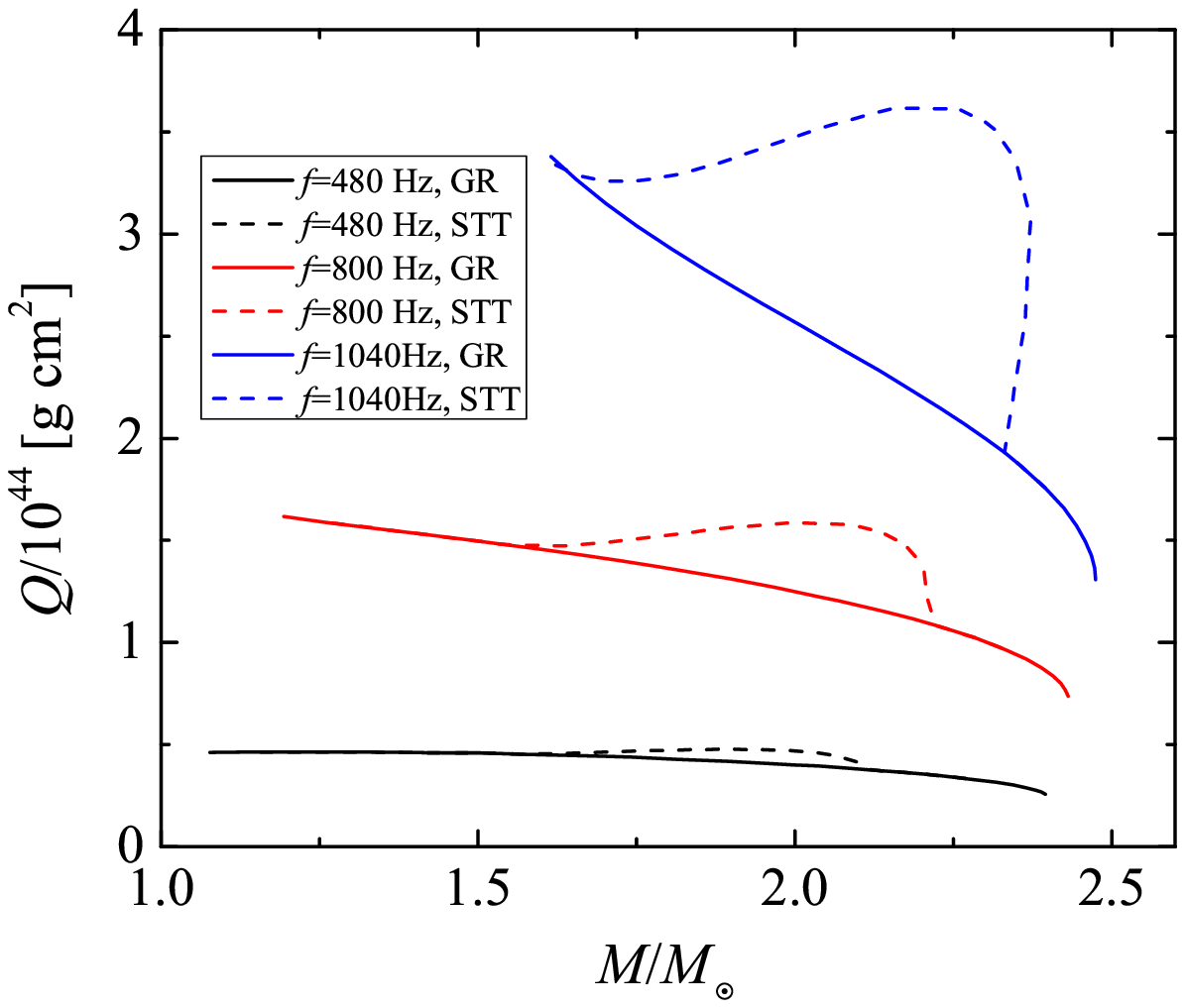}
\caption{The unnormalized moment of inertia (left panel) and quadrupole moment (right panel) as function of the neutron star mass for several sequences with fixed rotational frequency. The resented results are for EOS APR.} \label{Fig:IQ_NonNormalized}
\end{figure}

It is important to note that even though the normalized $\bar{I} - \bar{Q}$ relations are very close to GR for the case of $\beta=-4.5$, the unnormalized moment of inertia and quadrupole moment can deviate strongly from pure Einstein's theory. In Fig. \ref{Fig:IQ_NonNormalized} the unnormalized $I$ and $Q$ are shown as function of mass for several fixed values of the rotational frequency. It is evident that as the frequency increases the deviations from GR are magnified and the difference can reach much larger values compared to the normalized $\bar{I} - \bar{Q}$ relations. In the graph we have shown models with rotational periods up to approximately 1ms, but if one considers even faster rotation the deviations increase further, reaching close to the Kepler limit above $50\%$ for the moment of inertia and $100\%$ for the quadrupole moment.  This can potentially lead to observational manifestations.

\section{Conclusions}
In the present paper we investigated the normalized $\bar{I} - \bar{Q}$ relations for rapidly rotating neutron stars in scalar tensor theories of gravity. Our goal was to check the universality, both with respect to EOS and to the gravitational theory, that was observed for slow rotation \cite{Pani2014}. The work is motivated by the fact that rapid rotation can significantly influence the scalarization and increase the differences from GR \cite{Doneva2013}. Our results show the the EOS universality is well preserved but the rotation can magnify the deviation from GR. Still the differences are below roughly $5\%$ (for the current lower limit of the scalar-field coupling parameter $\beta=-4.5$) which is supposed to be smaller than the expected observational uncertainties.

We studied also the dependence of the $\bar{I} - \bar{Q}$ relations for rapidly rotating neutron stars on the coupling parameter $\beta$ and showed that as $\beta$ is decreased, the differences from GR increases. This is an expected effect as smaller negative $\beta$ lead to larger deviations from GR in the neutron star equilibrium properties. Still even for values of $\beta$ that are already ruled our by the observations the changes in the normalized $\bar{I} - \bar{Q}$ relations are not so strong.

An important observation is that although the normalized moment of inertia and quadrupole moment are very close to the pure general relativistic case, the unnormalized $I$ and $Q$ can change significantly due to the neutron star scalarization. This can also lead to distinct observational signatures. Another important conclusion that can be
drawn from our results is that although the normalized $\bar{I}-\bar{Q}$ relations have the very nice property of being independent from the EOS, they can suppress the deviations between general relativity and some alternative theories. That is why they alone are of limited use for testing Einstein's theory.

\section*{Acknowledgements}

We would like to thank George Pappas for  discussions and Emanuele Berti for critical reading of the manuscript and constructive suggestions. DD would like to thank the Alexander von Humboldt Foundation for a stipend. KK,  SY and KS would like to thank the Research Group Linkage Programme of the Alexander von Humboldt Foundation for the support. The support by
the Bulgarian National Science Fund under Grant DMU-03/6, by the Sofia University Research Fund under Grant
63/2014 and by the German Science Foundation (DFG) via SFB/TR7 is gratefully acknowledged. Partial support
comes from "New-CompStar", COST Action MP1304.

%%%%%%%%%%%%%%%%%%%%%%%%%%%%%%%%%%%%%%%%%%%%%%%%%%%%%%%%%%%%%%%%%%%%%%%%%%%%%%%

\bibliography{references}

\end{document}